\documentclass[12pt]{iopart}

\usepackage{times}

\usepackage{dsfont}
\usepackage{iopams}
\usepackage{graphicx}
\usepackage{bm}
\usepackage{bbm}
\usepackage{color}
\usepackage{verbatim}
\usepackage{tikz}
\usepackage{hyperref}
\usepackage{float}
\usepackage{harvard}

\begin{document}

\title{Driving protocol for a Floquet topological phase without static counterpart}

\author{A. Quelle$^1$, C. Weitenberg$^2$, K. Sengstock$^2$ and C. Morais Smith$^1$}
\address{$^1$ Institute for Theoretical Physics, Center for Extreme Matter and Emergent Phenomena, Utrecht University, Princetonplein 5, 3584 CC Utrecht, The Netherlands}
\address{$^2$ Institut f\"ur Laserphysik, Universit\"at Hamburg, Luruper Chaussee 149, 22761 Hamburg, Germany}

\begin{abstract}
Periodically driven systems play a prominent role in optical lattices. In these ultracold atomic systems, driving is used to create a variety of interesting behaviours, of which an important example is provided by topological states of matter. Such Floquet topological phases have a richer classification that their equilibrium counterparts. Although analogues of the equilibrium topological phases exist, which are characterised by a Chern number, the corresponding Hall conductivity, and protected edge states, there is an additional possibility. This is a phase that has vanishing Chern number and no Hall conductivity, but nevertheless hosts anomalous topological edge states.\cite{Rudner2013} Due to experimental difficulties associated with the observation of such a phase, it has not been experimentally realised so far. In this paper, we show that optical lattices prove to be a good candidate for both its realisation and subsequent observation, because they can be driven in a controlled manner. Specifically, we present a simple shaking protocol that serves to realise this special Floquet phase, discuss the specific properties that it has, and propose a method to experimentally detect this fascinating topological phase that has no counterpart in equilibrium systems.
\end{abstract}

\maketitle


\section{Introduction}

The field of optical lattices is a flourishing part of modern physics \cite{Bloch2008}. This is in large part due to the extreme tunability of ultracold atomic systems, which allows for the quantum simulation of many paradigmatic models in condensed matter. Since the experimental realisation of topological phases in condensed matter systems \cite{Konig2007,Hasan2010,Zhang2010}, there has been an intense activity to reproduce and manipulate such states in optical lattices \cite{Goldman2016b}. Important examples are the realization of lattices with artificial gauge fields \cite{Struck2012,Aidelsburger2013,Aidelsburger2015} and of topological band structures \cite{Wu2016}.

One possible way to implement the artificial gauge fields required to create such a system is by using Raman-assisted tunnelling \cite{Jaksch2003,Aidelsburger2011,Miyake2013,Mancini2015,Stuhl2015}, while an alternative is to shake the lattice periodically \cite{Eckardt2005,Lignier2007,Struck2011,Parker2013,Jotzu2014,Flaschner2016}. The effective stroboscopic Hamiltonian for such a periodically driven system is obtained using Floquet theory. In the high-frequency regime, the dynamics of the system can be described in terms of an effective static theory. However, due to the non-equilibrium nature of Floquet systems, a much richer behaviour is possible outside of the high-frequency regime \cite{Lindner2011,Ezawa2013,Fregoso2013,Wang2013,Carpentier2015,Quelle2014QSH,Kundu2014,Quelle2016NJOP}, and their topological classification is more complicated than that of equilibrium systems \cite{Nathan2015}: there is a state where all Chern numbers vanish, which is yet topologically non-trivial \cite{Kitagawa2010,Rudner2013,Reichl2014,Titum2016}. Because of the non-trivial topology, these systems host protected chiral edge modes, but there is no transverse conductivity in the bulk, due to the vanishing Chern number. As a consequence, the bulk is no longer robust against Anderson localisation, and it is possible to fully localise the bulk states while preserving the edge states \cite{Titum2016}. Although this behaviour is well understood from a theoretical viewpoint, and various models exhibiting these features have been studied \cite{Kitagawa2010,Rudner2013,Reichl2014,Titum2016}, this state has not yet been experimentally realised. 

In this paper, we propose a simple shaking protocol for a honeycomb optical lattice loaded with fermions that allows for the realisation of this exotic topological state, which bears no analogue in equilibrium systems. We construct the full topological phase diagram for the model, and determine which specific experimental parameters might be used to access the non-trivial phase with vanishing Chern number. Because the Hall conductivity vanishes in this system, the topological nature of this system must be determined by measuring the edge states directly, or by constructing the relevant topological invariant, which requires full tomography of the driving cycle. We show that a 2D honeycomb optical lattice for fermions has favourable properties for measuring the edge states directly.

The layout of the paper is as follows. Firstly, we review the relevant results about Floquet topological insulators (FTIs) in Sec.~\ref{Floquet}. More specifically, we discuss the topological classification of this system, which is necessary to construct the phase diagram. Then, we determine the time-dependent Hamiltonian corresponding to our shaking protocol in Sec.~\ref{Model}. In Sec.~\ref{Results}, we construct the phase diagram for the model, and provide the dispersion relation at characteristic values of the parameters, as well as an analysis of the robustness of the phase. Then, in Sec.~\ref{Experiments}, we discuss the conditions for an experimental observation of such a phase. Finally, we conclude in Sec.~\ref{Conclusion}.

\section{Floquet topological insulators}\label{Floquet}
Floquet theory applies to time-periodic Hamiltonians \cite{Sambe1972,Hemmerich2010}, for which the time-dependent Schr\"odinger equation has quasi-periodic solutions $\psi(t)=\exp\left(-i\epsilon t/\hbar\right)\phi(t)$, where $\phi$ is a periodic function in time and thus a solution of $H_F\phi(t)=\epsilon \phi(t).$ Here, the Floquet Hamiltonian is defined as 
\begin{eqnarray}\label{HF}
H_F:=-i\ln\left[U\right]/T,
\end{eqnarray}
where $U:=U(T,0)$ is the propagator from $t=0$ to $t=T$, i.e. over a single period. If $H_F$ exhibits topologically protected edge states for a finite system, one speaks of a FTI.  The propagator $U$ is unitary, so its spectrum lies on the unit circle in the complex plane. Because the logarithm maps the unit circle onto the real line, $H_F$ has real spectrum and is Hermitian. If one chooses the branch cut of the logarithm to lie in one of the band gaps of $U$, $H_F$ will have a top and a bottom energy band. However, if the gap containing the branch cut also contains gapless edge modes, these modes will connect the top and bottom bands of $H_F$ through the branch cut \cite{Rudner2013}. The full classification of a FTI, taking this effect into account, can be done in terms of winding numbers \cite{Rudner2013}, or Weyl cones \cite{Nathan2015}. The classification in terms of winding numbers is more closely analogous to the classification of equilibrium systems in terms of Chern numbers, so we recall it here. 

An $n$-band Floquet system has $n$ gaps $\Delta_i$, where $\Delta_i$ is the gap above band $i$. Note that due to the branch cut in $H_F$, $\Delta_0$=$\Delta_n$. One can associate the winding number $W_i$ to the gap $\Delta_i$ \cite{Rudner2013}. To construct $W_i$, note that the propagator is periodic in $\bm k$, but not necessarily in $t$, since $U=U(T,0)\neq \mathds{1}$ in general. First, we construct a periodic unitary operator by defining
\begin{equation}\label{PeriodicProp}
V_i(\bm k,t)=\left\{\begin{array}{cc}
      U\left(\bm k,2 t\right) & 0\leq t \leq \frac{T}{2} \\
      \exp\left(-i (2T-2t)H_F\right) & \frac{T}{2}\leq t \leq T
   \end{array}\right.
\end{equation}
This equation depends on a choice, since the eigenvalues of $U$ lie on the unit circle. One therefore has to choose a Brillouin zone for the quasi-energies of $H_F$. Mathematically, this corresponds to a choice of branch cut for the logarithm in Eq.~(\ref{HF}). This means that $V_i$ is only a continuous function of $\bm k,t$ if one chooses the ends of this Brillouin zone to lie in one of the band gaps $\Delta_i$ of $U$. In Eq.~(\ref{PeriodicProp}) we have assumed that these ends lie in the specific gap $\Delta_i,$ and we indicate this through the subscript of $V_i$.
The operator $V_i$ is periodic over the generalised Brillouin zone (GBZ) including time. Hence, we can define the winding number
\begin{eqnarray}\label{Winding}
W_i:=&\frac{1}{8\pi^2}\int_{\rm{GBZ}}dt dk_x dk_y
& \rm{Tr}\left(V_i^{-1} \partial_t V_i \left[V_i^{-1} \partial_{k_x} V_i,V_i^{-1}\partial_{k_y} V_i,\right]\right).
\end{eqnarray}
Because this integral is a topological winding number, and $V_i$ was defined to be continuous, it is quantised. As can be seen from Eq.~(\ref{Winding}), the winding number is also derived from the bulk states, but in contrast to the Chern number, it depends on the full time-evolution operator of the Floquet system, not only on its value at stroboscopic times. It can be shown that $W_i$ equals a sum over all chiral edge modes crossing $\Delta_i$, weighted with a $\pm$, depending on their propagation direction. The Chern number of band $i$ can be expressed as $C_i=W_i-W_{i-1}$, which yields the connection with the Chern number of the bands. In a static system, $W_n=W_0=0$, because the spectrum has no branch cut, but is bounded both below and above. This constraint allows one to express the $W_i$ in terms of $C_i$ uniquely. For a Floquet system, $W_0$ is non-zero when a chiral edge mode crosses the branch cut, and all Chern numbers vanish for a state with $W_i=c$ for all $i$, while it is topologically non-trivial if the integer $c$ is non-zero. In this case, it is the driving cycle that protects the edge modes, instead of the topology of the bands, as for conventional static systems. These trivial bands are not protected from Anderson localisation by topology, which allows for chiral edge modes in combination with a fully localised bulk \cite{Titum2016}.

\section{The Model}\label{Model}
It is known that a phase with $W_i=1$ can be created through periodic modulation of nearest-neighbour (NN) hopping parameters either in a bipartite square \cite{Rudner2013,Reichl2014} or in a hexagonal lattice \cite{Kitagawa2010}. In optical lattices, the hopping amplitudes in the different directions can be simply tuned by varying the intensities of the lattice beams \cite{Zhu2007}. However, we propose here a different approach using lattice shaking \cite{Koghee2012}, which has clear experimental advantages. First, the hopping amplitudes can be modulated more effectively, allowing for a full suppression of the undesired hopping amplitudes instead of just a finite anisotropy, thus realizing the model considered in Ref.~\cite{Kitagawa2010}. Second, the use of lattice shaking might allow for a cleaner implementation of the step-function-like switching of the hopping amplitudes that is at the origin of the perfectly flat bands we discuss later. Thirdly, lattice shaking couples differently to the higher bands than amplitude modulation of the lattice beams and therefore potentially leads to a smaller heating rate.

\begin{figure}
\centering\includegraphics[width=.5\textwidth]{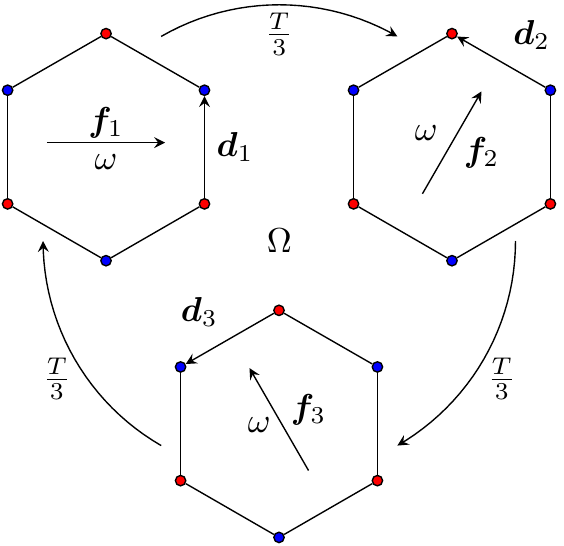}
\caption{\label{df} (Colour online) A scheme for the driving protocol is shown. The full driving cycle, with frequency $\Omega$, consists of three separate subcycles, which each last for time $T/3$, a third of the full driving period $T$. In each of these subcycles, the lattice is shaken in the direction of the vector $\bm f_i$ at frequency $\omega$. The shaking along $\bm f_i$ leaves the hopping parameter along the direction $\bm d_i$ unchanged, and may chosen such that it renormalises the hopping along the other two directions to zero. The bond vectors $d_i$ are also indicated in the figure, and the net result of the driving is that an electron hops counter-clockwise along the plaquette.}
\end{figure}

Since the model has two energy bands, we will label the gaps by their location in energy space for clarity. This allows us to define gaps $\Delta_0$ and $\Delta_{\Omega/2}$, with corresponding winding numbers. The driving protocol takes advantage of the fact that linear shaking of a lattice renormalises the hopping according to the projection of the shaking amplitude onto the bond \cite{Struck2011,Koghee2012}. Specifically, let the lattice be subjected to a sinusoidal shaking
\begin{equation*}
\bm F(t)=\sin(\omega t) \bm f,
\end{equation*}
where $\omega$ is the shaking frequency, and $\bm f$ a vector determining the shaking direction and amplitude, as depicted in Fig.~\ref{df}. Then, the NN hopping parameters $\gamma$ in a tight-binding model become renormalised as 
\begin{equation}\label{renormalisation}
\gamma\mapsto \gamma_r=\gamma J_0\left(\frac{m \omega \bm d\cdot \bm f}{\hbar}\right).
\end{equation}
Here, $\bm d$ is a vector in the bond direction with magnitude equal to the bond length, $m$ is the particle mass, and $J_0$ is the Bessel function of the zeroth kind. For future reference, let us now define $x_0$ to be the first zero of $J_0$. The expression in Eq.~(\ref{renormalisation}) is the first order in a perturbation expansion, and valid for large $\omega$. The tight-binding Hamiltonian then reads
\begin{eqnarray}\label{Ham}
H(\bm k)=
\gamma\sum_l\left( \begin{array}{cc}
0 & \exp(i {\bm k}\cdot \bm d_l)  \\
 \exp(-i {\bm k}\cdot \bm d_l) & 0
\end{array} \right).
\end{eqnarray}
Our convention for the NN hopping vectors $\bm d_l$ is $\bm d_1 = a(0,1),$ ${\bm d_2= a\left(-\sqrt{3},1\right)/2},$ and ${\bm d_3=- a\left(\sqrt{3},1\right)/2},$ where $a$ is the NN distance. By shaking in the 2D plane, perpendicularly to a particular bond, $\bm d\cdot \bm f=0$ for that bond, and $a f \cos(\pi/6)$ for the other two bonds. By choosing $f$ and $\omega$ in Eq.~(\ref{renormalisation}) such that 
\begin{eqnarray}\label{Scond}
m f a \omega\cos(\pi/6)/\hbar=x_0,
\end{eqnarray}
we can suppress two of the three hopping parameters through shaking, leaving the third one unaffected. This procedure yields the three renormalised Hamiltonians
\begin{eqnarray*}
H_l(\bm k)=
\gamma\left( \begin{array}{cc}
0 & \exp(i {\bm k}\cdot \bm d_l)  \\
 \exp(-i {\bm k}\cdot \bm d_l) & 0
\end{array} \right).
\end{eqnarray*}
Finally, we consider the system with Floquet propagator
\begin{eqnarray}\label{prop}
U(\bm k)=\prod_l\exp\left[-\frac{iT}{3\hbar}H_l(\bm k)\right],
\end{eqnarray}
where the product is ordered with higher indices to the left. It should be noted that we apply the Floquet theory two consecutive times, first to construct the effective Hamiltonians $H_l$, and then to obtain Eq.~(\ref{prop}). We have checked through a numerical calculation of $U$ that the errors induced by this perturbative treatment are negligible if one chooses $\omega$ large enough. The consecutive application of Floquet theory is only possible if $3\omega$ is a multiple of $\Omega:=2\pi/T$, since the three individual shaking protocols then fit in the driving cycle in a commensurate way, as depicted in Fig.~\ref{df}. The renormalised hopping in Eq.~(\ref{renormalisation}) imposes a constraint on $f \omega$, so by tuning the shaking amplitude $f$, one can achieve a commensurate $\omega$ at any desired value of $\gamma_r$.

\section{Results}\label{Results}
The Floquet propagator in Eq.~(\ref{prop}) can be calculated in closed form as long as translational symmetry is present in the system. To do so, let us write
\begin{eqnarray}
H_l(\bm k)=\gamma \left[\cos(\bm k\cdot \bm d_l)\sigma_x+\sin(\bm k\cdot \bm d_l)\sigma_y\right].
\end{eqnarray}
We can define partial propagators
\begin{eqnarray}
	U_l(\bm k):=\exp\left[-i \frac{T}{3\hbar} H_l\right]=\cos\left(\frac{\gamma T}{3\hbar}\right)-\frac{i}{\gamma}\sin\left(\frac{\gamma T}{3\hbar}\right)H_l(\bm k).
\end{eqnarray}
Here we have made use of the fact that all the eigenvalues of $H_l$ are $\pm \gamma$, independent of $\bm k$, so that $H_l(\bm k)/\gamma$ has unit norm on the Bloch sphere. It follows from the definition that $U=U_3 U_2 U_1$. To evaluate the various cross-terms in this product, we use
\begin{eqnarray*}
\frac{1}{\gamma^2}H_m(\bm k) H_n(\bm k)=\cos\left[\bm k\cdot(\bm d_m-\bm d_n)\right]+i \sin\left[\bm k\cdot(\bm d_m-\bm d_n)\right]\sigma_z,\\
\qquad \frac{1}{\gamma^3}H_3(\bm k) H_2(\bm k) H_1(\bm k)=\sigma_x.
\end{eqnarray*}
The last equality follows because in our convention $\bm d_1+\bm d_3=\bm d_2$, which simplifies the product. This allows us to write
\begin{eqnarray}\label{FloqProp}
U(\bm k)=\mathds{1}\left[\cos^3(\tau)-\cos(\tau)\sin^2(\tau)\right.\\
\left.\left\{\cos\left[\bm k\cdot(\bm d_1-\bm d_2)\right]+\cos\left[\bm k\cdot(\bm d_1-\bm d_3)\right]+\cos\left[\bm k\cdot(\bm d_2-\bm d_3)\right]\right\}\right]\nonumber\\
+i\sigma_x\left\{\sin^3(\tau)-\cos^2(\tau)\sin(\tau)\left[\cos\left(\bm k\cdot\bm d_1\right)+\cos\left(\bm k\cdot\bm d_2\right)+\cos\left(\bm k\cdot \bm d_3\right)\right]\right\} \nonumber\\
-i\sigma_y\cos^2(\tau)\sin(\tau)\left[\sin\left(\bm k\cdot\bm d_1\right)+\sin\left(\bm k\cdot\bm d_2\right)+\sin\left(\bm k\cdot \bm d_3\right)\right]\nonumber\\
-i\sigma_z\left[\cos(\tau)\sin^2(\tau)\right.\nonumber\\
\left.\left\{\sin\left[\bm k\cdot(\bm d_1-\bm d_2)\right]+\sin\left[\bm k\cdot(\bm d_1-\bm d_3)\right]+\sin\left[\bm k\cdot(\bm d_2-\bm d_3)\right]\right\}\right].\nonumber
\end{eqnarray}

\begin{figure}
\centering\includegraphics[width=.5\textwidth]{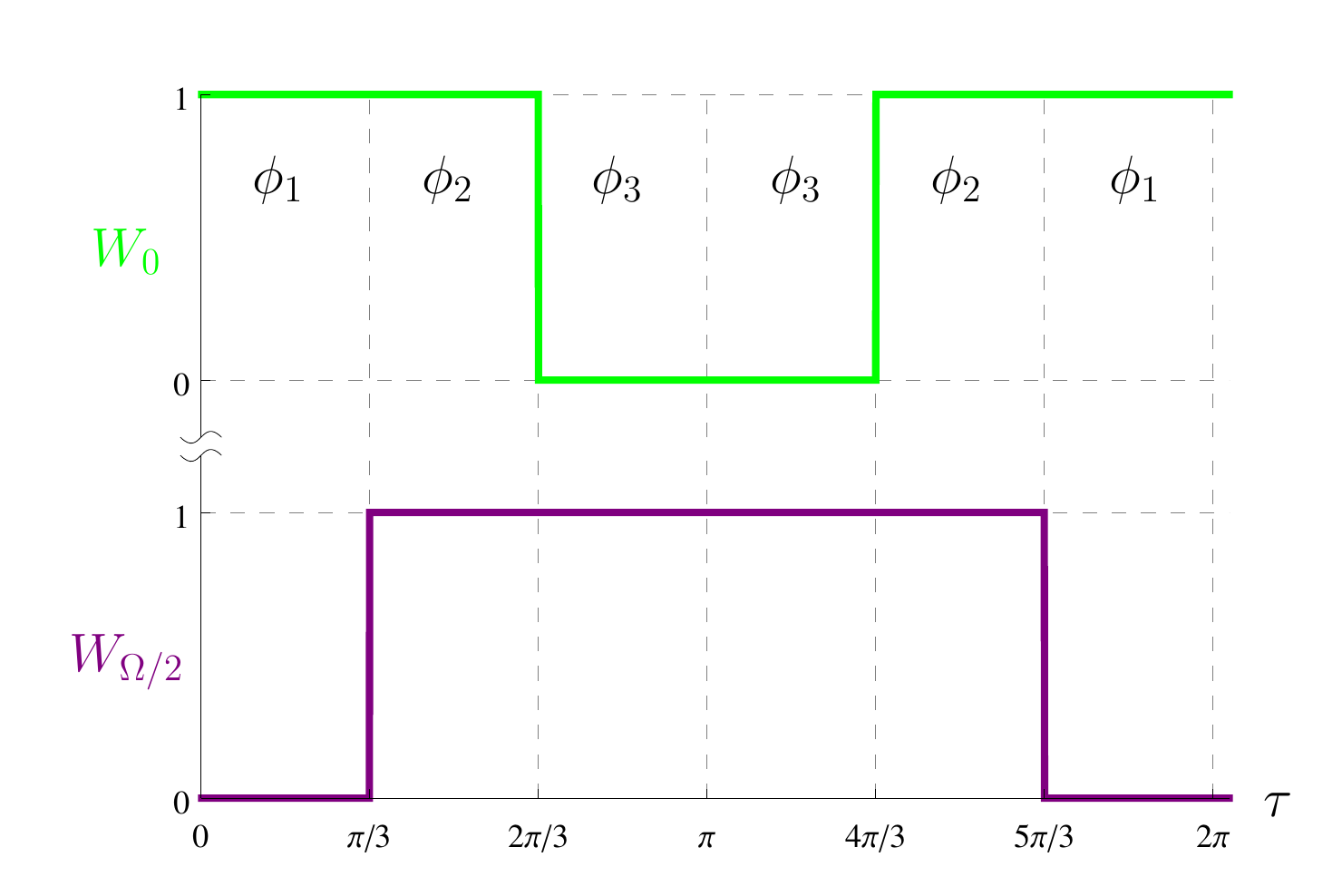}
\caption{\label{pd} (Colour online) The winding numbers $W_0$ and $W_{\Omega/2}$ are plotted in green and purple, respectively. The different combinations of these two numbers yield three different phases, which are indicated by $\phi_i$, for $i=1,2,3$. It should be noted that the gap closes at $\tau=\pi$, so $\phi_3$ is not well defined at this point; the gap closing does not change the topological phase, however. The Chern numbers of the bands are given by the difference of the two winding numbers, and therefore take the values $0$ and $\pm 1$.}
\end{figure}

\begin{figure}[h!]
\centering\includegraphics[width=.5\textwidth]{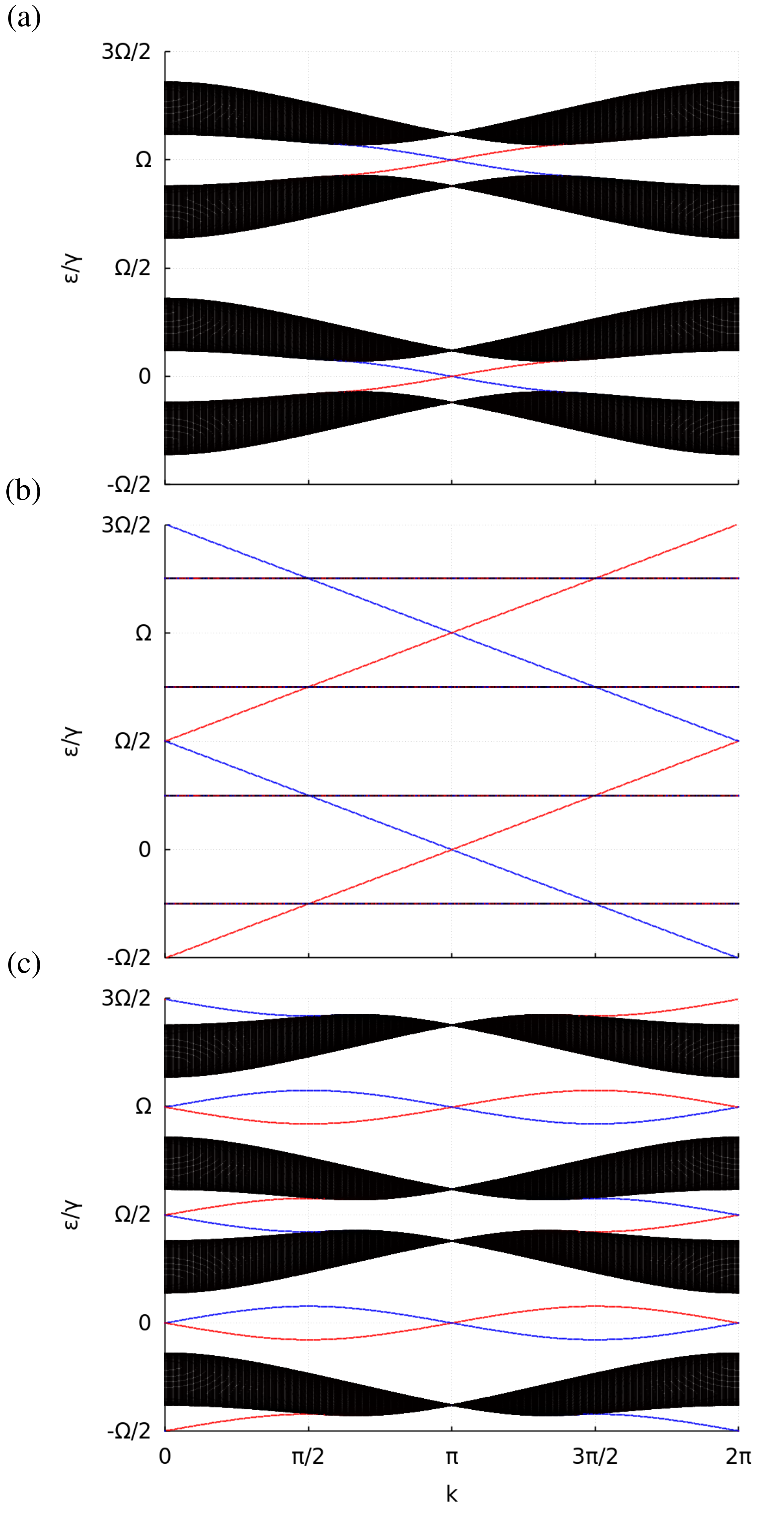}
\caption{\label{disp} (Colour online) The dispersion of the propagator in Eq.~(\ref{prop}) is shown for a ribbon geometry with zigzag edges for three different values of the driving parameter $\tau$. In all cases there are two energy bands indicated in black, separated by gaps $\Delta_0$ and $\Delta_{\Omega/2}$. The blue and red states are localised on the top and bottom edge, respectively. (a) The parameter $\tau=6\pi/25$, so the system is in phase $\phi_1$. For smaller $\tau$, i.e. larger $\Omega$, the dispersion looks similar, but the gap around $\Omega/2$ becomes larger, so the system behaves more like a static one. (b) The parameter $\tau=\pi/2$, so the system is in phase $\phi_2$. By tuning the frequency away from this point, the bulk bands will no longer be flat, but the topological behaviour will be the same. (c) The parameter $\tau=19\pi/25$, so the system is in phase $\phi_3$. The dispersion is like a flipped version of that in (a), where the two gaps have been interchanged. The only difference is the presence of trivial edge states in $\Delta_0$.}
\end{figure}

Here, $\tau=T\gamma/3\hbar$ is the dimensionless parameter characterising the driving frequency. The full topological phase diagram can only be obtained by considering the driving cycle leading to this operator through Eq.~(\ref{Winding}). However, some information can already be gleaned from Eq.~(\ref{FloqProp}). For example, sending $\tau\mapsto \tau+\pi$ sends $U\mapsto-U$, and consequently $H_F\mapsto H_F+\omega/2$. This shows that the bands are exchanged under this transformation, and so are their Chern numbers. The method above can also be used to calculate $U(t,0)$, from which one can determine $W_0$ and $W_{\Omega/2}$ to construct the phase diagram. This phase diagram is $2\pi$ periodic in $\tau$, and the first period is depicted in Fig.~\ref{pd}. Three phases are visible, all of which are topological due to the presence of a gapless edge mode in at least one gap. In the following, we discuss the phases $\phi_i$, $i=1,2,3$. Representative dispersions from these phases are plotted in Figs.~\ref{disp}(a),(b),(c), respectively, for a ribbon geometry with zigzag edges. 

In Fig.~\ref{disp}(a), we have plotted the dispersion of $H_F$, as defined using Eqs.~(\ref{HF}) and \ref{prop}, for the parameter value $\tau=6\pi/25.$ The spectrum is shown for two periods of the quasi-energy, to illustrate that $W_0=1$ (as evidenced by the chiral edge mode), while $W_{\Omega/2}=0$. Consequently, this phase also has non-vanishing Chern number. By increasing the frequency, the gap around $\Omega/2$ increases in size (because the period of the spectrum increases, but the bandwidth does not), and one reaches a high-frequency regime in which the system is well described by a static Hamiltonian. The phase of this effective static Hamiltonian, with a single edge state between the two bands, is the topological phase of the Haldane model \cite{haldane,Quelle2016NJOP}.

A situation that is only possible for Floquet systems occurs when one increases $\tau$, i.e. if one lowers the frequency. When the frequency becomes low enough, the energy bands from different periods of the quasi-energy start to overlap, a situation that physically corresponds to the appearance of resonances due to the driving \cite{Quelle2016NJOP}. In this case, one enters the phase $\phi_2$, where these driving resonances also cause topological edge states to appear in the gap around $\Omega/2$, as depicted in Fig.~\ref{disp}(b).

The distinguishing feature of phase $\phi_2$ is that $W_0=W_{\Omega/2}=1$, meaning that the Chern numbers of both bulk bands vanish, while each gap hosts an edge mode. For the specific value of $\tau$ used in Fig.~\ref{disp}(b), the bulk bands of the system are also dispersionless, an interesting feature, since the flat bands together with vanishing Chern number imply totally localised bulk electrons. Note that in this case, the localisation is not due to Anderson localisation, but due to the driving protocol, making it a different situation from that in Ref.~\cite{Titum2016}. The non-zero winding numbers associated to the gap are induced by the chiral nature of the driving protocol, and are thus topological. Consequently, this state can only occur in Floquet systems.

The final phase $\phi_3$ is obtained by increasing $\tau$ further. The appearance of a two photon resonance in the system destroys the topological protection of the edge state in $\Delta_0$, as can be seen in Fig.~\ref{pd}, which shows that $W_{\Omega/2}=1$, while $W_0=0$. A respresentative dispersion relation is shown in Fig.~\ref{disp}(c). The winding numbers for $\phi_3$ imply that the edge state in $\Delta_0$ is not protected by topology, which is consistent with the fact that it is not chiral.

Being topological, these phases should be robust against various kinds of disorder. Due to our intended application in optical lattices, which are inherently defect free, we omit a detailed discussion of the influence of disorder. It is expected that lattice disorder will not destroy the topological phase if the disorder is small enough compared to the gap size. Of greater interest is the robustness of the phase to the parameters that depend on the driving protocol: the Floquet propagator in Eq.~(\ref{prop}) is built from effective Floquet Hamiltonians where only one of the NN hopping parameters in the honeycomb lattice is assumed to be non-zero. In general, this will not be precisely true, and one of the hopping parameters will merely be much larger than the others. There are two possibilities: the larger hopping parameter may have the same sign as the two smaller ones, or the opposite sign. In both cases, the qualitative behaviour is the same, as we describe in the following. Firstly, it should be noted that the driving frequencies at which the winding numbers $W_i$ change will differ slightly from those depicted in Fig.~\ref{pd}. Nevertheless, the phase $\phi_2$ from Fig.~\ref{disp}(b) persists even if the smaller hopping parameters become as large as $\gamma/10$, if one is at the point $\tau=\pi/2$. The larger $\tau$ becomes, the larger are the deviations from the ideal case presented in Eq.~(\ref{prop}), so the allowed uncertainty in the hopping parameters depends on the value of $\tau$ that one intends to work with. 

A similar discussion can be held with respect to the presence of next-nearest-neighbour (NNN) hopping. Since NNN hopping naturally occurs in optical lattices, a treatment of its effects is important to connect with experiments. In the honeycomb lattice, there are six NNN hopping vectors. These have length $\sqrt{3} a$, where $a$ is the NN bond length, as defined previously. Two of these are perpendicular to $\bm d_1$, two to $\bm d_2$, and two to $\bm d_3$. When shaking according to the protocol discussed above, four of the NNN hopping parameters get renormalised to zero, and the two parallel to the shaking pick up a factor $J_0(2 x_0)\approx -0.24$. Hence, the shaking protocol has the added benefit of strongly suppressing the NNN hopping contribution. Now, the phase $\phi_2$ is accessible when the {\it renormalised} NNN hopping parameter is smaller than approximately $\gamma/4$. The {\it bare} NNN hopping strength is dependent on the lattice depth, and below we will consider a value of $4\gamma/100$ \cite{Ibanez2013}. This is clearly within the required range, so NNN hopping will not influence the experimental realisability of this phase. It must be noted that in the presence of NNN hopping, the bulk bands are no longer completely flat, but the topological characteristics of the phase remain unchanged.

\section{Experimental Realisation}\label{Experiments}

It should, therefore, be possible to tune the hopping parameters in such a way that the phase $\phi_2$, which is characterised by a non-trivial topological structure but vanishing Chern numbers, can be reached. In light of this fact, we will now discuss some possible experimental parameters that might allow the experimental realisation of the phase $\phi_2$ in optical lattices. 

The condition that the renormalised hopping parameters in Eq.~(\ref{renormalisation}) vanish imposes a constraint on the shaking amplitude and frequency: $J_0\left(m f a \omega\cos(\pi/6)/\hbar\right)=0$. Assuming that one uses the first solution to this equation, one can rewrite it in terms of the recoil energy $\hbar \pi^2/2m a^2$ as
\begin{eqnarray}\label{constraint}
\frac{\omega}{\omega_{rec}}=\frac{4 x_0}{\sqrt{3} \pi^2}\frac{a}{f}.
\end{eqnarray}
The recoil energy depends on the particle mass and the lattice constant, and it is the only parameter in Eq.~(\ref{constraint}) that depends on the atomic species loaded into the lattice. For the realisation of the phase $\phi_2$, taking $\tau=\pi/2 \rm{mod} \pi$ is preferable. As shown in Fig.~\ref{disp}(b), the bands are flattest for this value, which is desirable for the reasons that we discuss below. The highest total frequency for which this holds is $\hbar\Omega=4 \gamma/3.$ From now on, we will assume that $\hbar \Omega$ takes this maximal value, since it corresponds to the shortest time scale for the experiment. Since the two shaking frequencies have to be commensurate, $\omega=3 n \Omega$ for $n\in \mathbb{N}$, which corresponds to a shaking amplitude $f$ given by Eq.~(\ref{constraint}). Because of Eq.~(\ref{constraint}), increasing $n$ requires a lowering of $f$, but since both parameters can be tuned within a wide range, many possible values can be chosen. For concreteness, we will assume $n=2$, which gives the minimum shaking frequency where our model is accurate, and which minimises resonant coupling of the system to higher bands. The NN hopping of fermions in an optical lattice is usually expressed in terms of $\omega_{rec}$ since it naturally incorporates the effect of particle mass and the lattice constant. We consider a lattice depth of $7 \omega_{rec}$, which corresponds to $\gamma=\hbar\omega_{rec}/10$, and a NNN hopping of $4 \omega_{rec}/1000$ \cite{Ibanez2013}. Using these values for the hopping parameters, and combining these with the chosen value of $\tau$, we find the driving frequency of the system to be $\Omega\approx 4 \omega_{rec}/30$, which is about $0.13\omega_{rec}$.

We can obtain specific numbers by choosing a particle mass and an optical wavelength, which allows us to specify $a$ and $\omega_{rec}$. Let us consider fermionic $^{40}$K loaded in a honeycomb optical lattice with wavelength $\lambda=1064 \rm{nm}$, which amounts to a recoil frequency $\omega_{rec}/2\pi=4.41 \rm{kHz}$, and consequently, the minimal driving period $T=1.7 \rm{ms}$. The minimal commensurate shaking frequency $\omega$ is then $0.4 \omega_{rec}/2\pi\approx 1.76 \rm{kHz}$, which corresponds to a maximal shaking amplitude of $0.075 a$. Since the lattice constant is $2\lambda/3$ for a honeycomb lattice, the maximum shaking amplitude $f\approx 53 \rm{nm}.$

As mentioned earlier, the same phase can be obtained for a variety of parameters. For instance, the frequency $\omega$ can be chosen to be any multiple of $0.4 \omega_{rec}$, as might be desired to minimise coupling to other bands in the lattice. The shaking amplitude will then be the corresponding fraction of $53 \rm{nm}$. Furthermore, due to the periodicity of the phase diagram, the phase $\phi_2$ can also be realised most generally if $(2n+1)\hbar\Omega=4 \gamma/3.$ This allows one to shake at lower frequencies $\omega$, at the cost of lower $\Omega$, which might be disadvantageous due to heating of the system.

The distinguishing feature of the phase $\phi_2$ is the presence of edge states, while there is neither time-reversal symmetry, nor a non-vanishing Chern number. The Chern number can be measured in terms of the Hall conductivity, which has been achieved in optical lattices \cite{Aidelsburger2015}, so one can experimentally prove that the Chern number vanishes in this topological phase. There are, nevertheless, edge modes present in the system, protected by the topological nature of the winding numbers associated to the quasi-energy gaps in the system. 

It is well known that the Hall conductivity is quantised in terms of the Chern number. The fact that the phase $\phi_2$ is dynamic in origin does not alter this conclusion. The total wavefunction corresponding to a single occupied electron band of the Floquet propagator is time dependent, returning to itself only after each period $T$, but this time evolution is unitary. The Chern number, because it is a topological invariant, is invariant under unitary transformations, so the Hall conductivity is constantly zero.

It is, therefore, necessary to detect the topological phase in a different way. One alternative would be to directly detect the edge states to prove the presence of a topological phase in the system. The principal difficulty in 2D systems is the presence of dispersive bulk bands, together with the fact that the edge mode is not present for all $k$. For this reason, an electron inserted at the edge of a system will have an overlap with the bulk modes, since a localised electron has equal overlap with all momenta. If the bulk is dispersive, this overlap will cause the electron wavefunction to partially leak away into the bulk, making a measurement of the edge state difficult at longer timescales. As can be seen in fig.~\ref{disp}(b), for certain parameter values the bulk is nearly dispersionless, while the edge states exist for nearly all $k$-values. This makes the currently proposed system conducive to the direct measurement of topological edge states. 

\begin{figure}
\centering\includegraphics[width=.5\textwidth]{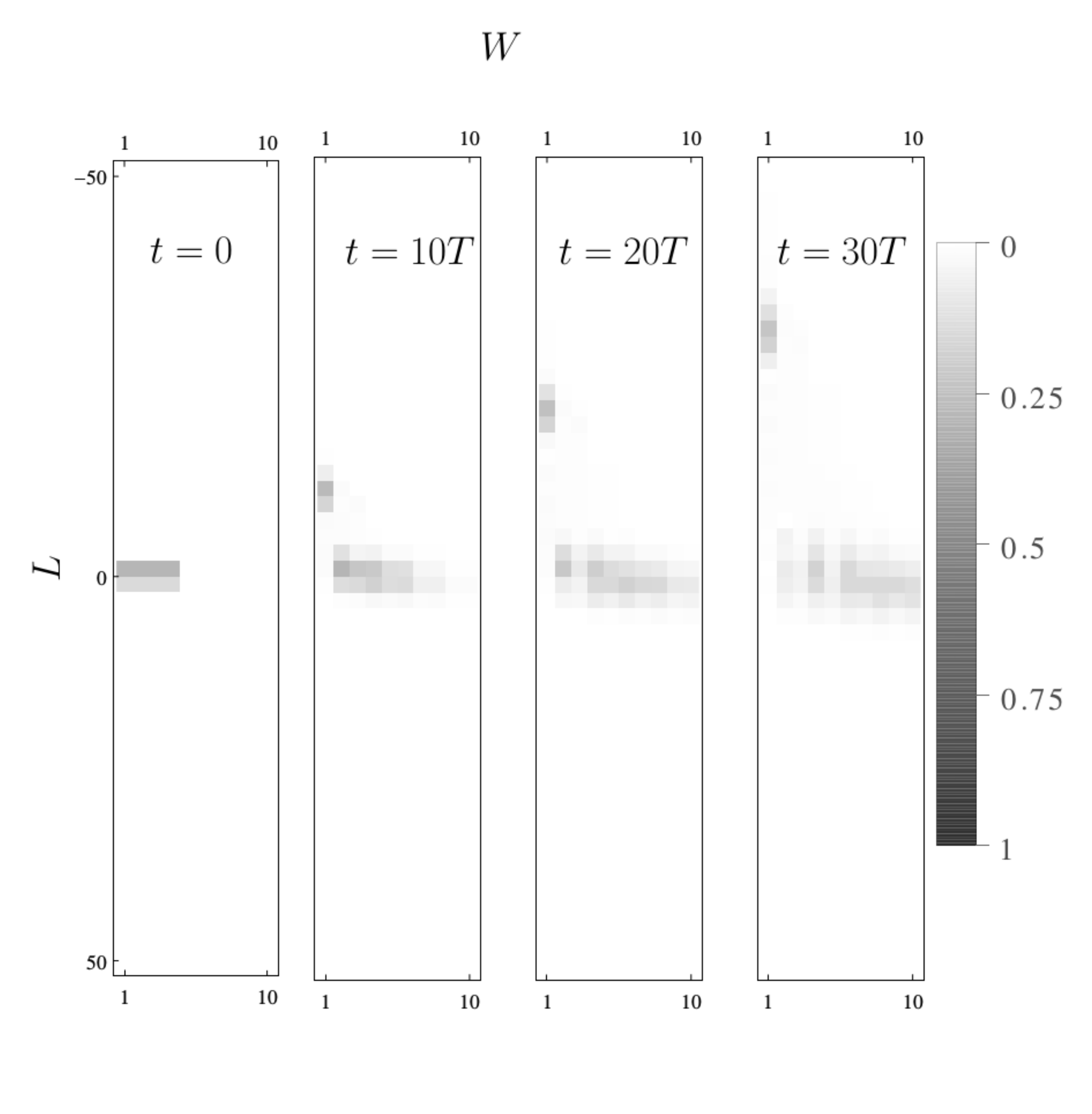}
\caption{\label{wf} The probability density is depicted at 4 different times for a cylindrical system with zigzag edges, for $\tau=\pi/2$ and a {\it bare} NNN hopping of $\gamma/10,$ which corresponds to the parameter values discussed in the text. The coordinates $L,W$ label sites of the lattice using the lattice vectors. The lattice vector associated with $L$ lies parallel to the edge, meaning that $L$ is periodic, while the vector associated to $W$ points inwards. It should be noted that $W$ is also used to label sublattice, so that $W=1,2$ lie on the edge, and $W=3,4$ lie one unit cell inward, etc. The parameter ranges are $L\in (-200,200]$ and $W\in[1,500]$ (meaning $250$ unit cells), but due to the localisation of the wavefunction, only part of the system is shown.}
\end{figure}

To make the behaviour of edge states in this system explicit, we have shown the explicit time evolution of one such state in Fig.~\ref{wf}. Specifically, we have plotted the wavefunction at different times to show its localisation on the edge. The initial state was spread with constant amplitude over block of $3$ unit cells along the boundary by $2$ unit cells inwards. The time-evolution indicates that the part of the wavefunction located on the $A$ sites of the edge (the tips of the zigzag, at $W=1$) stay localised there with a probability of over $90\%$, while the part of the wavefunction at $W>1$ slowly disperses into the bulk. Because the bulk is only weakly dispersive, the edge states have a higher group velocity, and the edge part of the wavefunction clearly separates from the bulk part. This shows that the combination of very flat bulk bands with edge states that exist for most $k$ values ensures that the edge and bulk parts of a wavefunction can be clearly separated from the dynamics. It should be noted that this discussion holds for an infinitely steep edge potential.


Instead of detecting the edge state directly, an alternative experimental approach would be to obtain the bulk winding number in Eq.~(\ref{Winding}), which dictates the presence of edge states. The newly developed state tomography \cite{Flaschner2016} yields full access to the time-dependent bulk Bloch states. By reconstructing both the Chern number of the effective Floquet Hamiltonian obtained from the stroboscopic time steps and the winding number obtained from the full time-dependent state, one can experimentally disentangle the two topological indices.

\section{Conclusion}\label{Conclusion}

Floquet systems allow for the realisation of a curious topological phase, which is characterised by vanishing Chern numbers for all bands, but exhibits topologically protected edge states. Due to the vanishing Chern number, there is no Hall conductivity in the bulk, and the bulk electrons can localise. This has a surprising consequence for Laughlin's charge pumping argument \cite{Laughlin1981}. Since there is no Hall conductivity present in the system, threading a flux through it, or briefly turning on an electric field, will not pump charge across the system. This can be seen from the spectrum in Fig.~\ref{disp}(b): the topological edge modes form a closed loop, which is only possible due to the branch cut. Threading a flux through the system merely moves electrons through the loop, but they never move into the bulk. 

Several models that exhibit such a Floquet topological phase have been proposed \cite{Kitagawa2010,Rudner2013,Reichl2014,Titum2016}, and an experimental realisation would be desirable. We propose a simple shaking protocol for a honeycomb optical lattice that allows for an experimental realisation of the model in Ref.~\cite{Kitagawa2010}, and we discuss possible experimental advantages of this approach. Because Chern numbers have been measured in 2D Floquet optical lattices \cite{Jotzu2014,Aidelsburger2015,Flaschner2016}, it is experimentally possible to show the vanishing of the Hall conductivity in this phase. The direct detection of the topological edge states remains an experimental challenge: so far, edge states have only been experimentally observed in 1D systems \cite{Leder2016}, ladder systems \cite{Atala2014,Tai2016} or artificial dimensions \cite{Mancini2015,Stuhl2015}. However, promising proposals for their detection in 2D systems exist using either Raman spectroscopy \cite{Goldman2012b} or the different dynamics of the bulk and edge states after a removal of a barrier \cite{Goldman2013}. 
Detection methods involving sharp walls are especially promising for the system under discussion. By tuning the parameters, it is possible to make the bulk bands nearly flat, ensuring that the wavefunction of an electron injected at such a wall has minimal leakage into the bulk.

Beyond the scenario of edges induced by sharp walls, a promising direction are interfaces between regions of different topology induced by spatially varying lattice parameters \cite{Reichl2014,Goldman2016}. This scenario could apply to our proposal because the phase transitions are controlled by the shaking frequency relative to a resonance that depends on the spatially varying lattice depth.  

Alternatively, recent advances in state tomography \cite{Flaschner2016} could allow for a direct measurement of the winding number in Eq.~(\ref{Winding}). Together, these properties might make the present proposal a promising candidate for the experimental realisation and detection of a Floquet topological phase that has no static counterpart.

The work by A.Q. and C.M.S. is part of the D-ITP consortium, a program of the Netherlands Organisation for Scientific Research (NWO) that is funded by the Dutch Ministry of Education, Culture and Science (OCW). C.W. and K.S. acknowledge financial support from the excellence cluster "The Hamburg Centre for Ultrafast Imaging - Structure, Dynamics and Control of Matter at the Atomic Scale".\\

\bibliographystyle{jphysicsB}

\end{document}